\documentclass[12pt,technote,onecolumn,doublespace]{IEEEtran}
\usepackage{amssymb}
\usepackage{amsmath}

\begin{document}

\title{Periodic Chandrasekhar recursions}
\author{Abdelhakim~Aknouche* and Fay\c{c}al Hamdi*\thanks{%
*Department of Operation Research, Faculty of Mathematics, University of
Sciences and Technology Houari Boumediene, Algiers, Algeria.}}
\maketitle

\begin{center}
\textbf{Abstract}
\end{center}

This paper extends the Chandrasekhar-type recursions due to Morf, Sidhu, and
Kailath "Some new algorithms for recursive estimation in constant, linear,
discrete-time systems, \textit{IEEE Trans}.\textit{\ Autom}.\textit{\ Control%
} 19 (1974) 315-323" to the case of periodic time-varying state-space
models. We show that the $S$-lagged increments of the one-step prediction
error covariance satisfy certain recursions from which we derive some
algorithms for linear least squares estimation for periodic state-space
models. The proposed recursions may have potential computational advantages
over the Kalman Filter and, in particular, the periodic Riccati difference
equation.

\noindent \textbf{Keywords}: Periodic state-space models, Chandrasekhar-type
recursions, Kalman Filter, periodic Riccati difference equation.

\section*{Introduction}

Morf et \textit{al} $\left( 1974\right) $ proposed recursions that
substitute the Kalman Filter for linear least squares estimation of
discrete-time time-invariant state space models, with a simpler
computational complexity. The new algorithms have been called
Chandrasekhar-type recursions because they are analog to certain
differential equations encountered in continuous-time problems (Kailath, $%
1973$). Since there, a considerable attention has been paid in the three
recent decades to the Chandrasekhar-type recursions (see e.g. Friedlander et
\textit{al}, $1978$; Morf and Kailath; $1975$; Houacine and Demonent, $1986$%
; Houacine, $1991$; Nakamori et \textit{al}, $2004$; Nakamori, $2007$). At
present, there exist several useful applications of the Chandrasekhar filter
in improving computational aspects related to the building of linear
time-invariant models. We mention non exhaustively the likelihood evaluation
(see Pearlman, $1980$; M\'{e}lard, $1984$; Kohn and Ansley, $1985$ for
autoregressive moving average, $ARMA$, models and Shea, $1989$ for vector $%
ARMA$ models), the calculation of the exact Fisher information matrix (see M%
\'{e}lard and Klein $(1994)$ for the $ARMA$ case and Klein et \textit{al} $%
(1998)$ for general dynamic time-invariant models), and the development of
fast variants of the recursive least squares algorithm (Houacine, $1991$;
Sayed and Kailath, $1994$; Nakamori et \textit{al} $2004$). As is well
known, the Chandrasekhar equations are restricted to the case of
time-invariant state-space models because of their particular time
invariance structure and it seems that there is no results tied to the class
of all nonstationarity, except in very special cases (Sayed and Kailath, $%
1994$). A particular class of nonstationarity whose importance has no need
to be proven is the one of periodic linear models. Important progress has
been made recently in the building and analysis of periodic $ARMA$ ($PARMA$)
and periodic state-space characterizations. The objective was to develop
extensions of similar methods for standard time-invariant models to their
periodic counterparts, without transforming periodic systems to their
corresponding multivariate time-invariant representations in order to
simplify the computational burden. Despite the current abundance of
computational methods for periodic state-space models (see e.g. Lund and
Basawa, $2000$; Varga and Van Dooren, $2001$; Gautier, $2005$; Bentarzi and
Aknouche, $2005$; Aknouche, $2007$; Aknouche and Hamdi, $2007$ Aknouche et
\textit{al}, $2007$ and the references therein) it seems that there is no
results concerning extensions of the Chandrasekhar recursions to the
periodic case. This paper proposes some algorithms for linear least squares
estimation of periodic state-space models. Our methods extend the
Chandrasekhar algorithms proposed by Morf et \textit{al} $(1974)$ to the
periodic time-varying case and retain their desirable features. As a result,
the periodic Chandrasekhar recursions are used through the innovation
approach to efficiently evaluate the likelihood of periodic $ARMA$ models.

The rest of this paper is organized as follows. Section 1 briefly recalls
some preliminary definitions and facts about periodic state-space models and
their corresponding Kalman Filter. In Section $2$ we develop some
Chandrasekhar-type algorithms that substitute the Kalman filter for periodic
state-space models. The initialization problem will be studied in Section 3.

\section{Preliminary definitions and notations}

Consider the following linear periodic state-space model%
\begin{equation}
\left\{
\begin{array}{l}
\mathbf{x}_{t+1}=F_{t}\mathbf{x}_{t}+G_{t}\mathbf{\epsilon }_{t} \\
\mathbf{y}_{t}=H_{t}^{\prime }\mathbf{x}_{t}+\mathbf{e}_{t}%
\end{array}%
\right. \text{, }t\in \mathbb{Z},  \tag{$1$}
\end{equation}%
where $\left\{ \mathbf{x}_{t}\right\} $,\ $\left\{ \mathbf{y}_{t}\right\} $,
$\left\{ \mathbf{e}_{t}\right\} $ and $\left\{ \mathbf{\epsilon }%
_{t}\right\} $\ are random processes of dimensions $r\times 1$, $m\times 1$,
$m\times 1$, and $d\times 1$ respectively, with%
\begin{equation*}
\left\{
\begin{array}{l}
E\left( \mathbf{\epsilon }_{t}\right) =E\left( \mathbf{e}_{t}\right) =0 \\
E\left( \mathbf{\epsilon }_{t}\mathbf{\epsilon }_{t+h}^{\prime }\right)
=\delta _{h,0}Q_{t} \\
E\left( \mathbf{e}_{t}\mathbf{e}_{t+h}^{\prime }\right) =\delta _{h,0}R_{t}%
\end{array}%
\right. \text{ and }\left\{
\begin{array}{l}
E\left( \mathbf{\epsilon }_{t}\mathbf{x}_{t-k}^{\prime }\right) =0 \\
E\left( \mathbf{e}_{t}\mathbf{y}_{t-k}^{\prime }\right) =0 \\
E\left( \mathbf{x}_{t}\mathbf{x}_{t}^{\prime }\right) =W_{t}%
\end{array}%
\right. ,\text{ }%
\begin{array}{c}
\forall t,h\in \mathbb{Z} \\
\forall k\geq 0%
\end{array}%
,
\end{equation*}%
($\delta $ stands for the Kronecker function). The nonrandom matrices $F_{t}$%
,\ $G_{t}$, $H_{t}^{\prime }$, $Q_{t}$, $R_{t}$,\ and $W_{t}$ are periodic
in time with period $S$. To simplify the exposition we suppose without loss
of generality that%
\begin{equation*}
E\left( \mathbf{e}_{t}\mathbf{\epsilon }_{l}^{\prime }\right) =0,\ \forall
t,l\in \mathbb{Z}.
\end{equation*}%
Let $\widehat{\mathbf{x}}_{t}$ and $\widehat{\mathbf{y}}_{t}$ be the linear
least squares forecasts of$\ \mathbf{x}_{t}$ and $\mathbf{y}_{t}$,
respectively, based on $\mathbf{x}_{1},\mathbf{x}_{2}...,\mathbf{x}_{t-1}$.
Then as is well known, $\widehat{\mathbf{x}}_{t}$\ and $\widehat{\mathbf{y}}%
_{t}$\ may be uniquely obtained from the Kalman filter (Kalman, $1960$)
which is given by the following recursions%
\begin{equation}
\left\{
\begin{array}{l}
\left( a\right) \quad \Omega _{t}=H_{t}^{^{\prime }}\Sigma _{t}H_{t}+R_{t},
\\
\left( b\right) \quad K_{t}=F_{t}\Sigma _{t}H_{t}, \\
\left( c\right) \quad \widehat{y}_{t}=H_{t}^{^{\prime }}\widehat{x}_{t}, \\
\left( d\right) \quad \widehat{x}_{t+1}=\left( F_{t}-K_{t}H_{t}^{^{\prime
}}\right) \widehat{x}_{t}+K_{t}y_{t}, \\
\left( e\right) \quad \Sigma _{t+1}=F_{t}\Sigma _{t}F_{t}^{\prime
}-K_{t}\Omega _{t}^{-1}K_{t}^{^{\prime }}+G_{t}Q_{t}G_{t}^{\prime },%
\end{array}%
\right.  \tag{$2$}
\end{equation}%
with starting values%
\begin{equation*}
\left\{
\begin{array}{l}
\left( f\right) \quad \widehat{\mathbf{x}}_{1}=E\left( \mathbf{x}_{1}\right)
=0, \\
\left( g\right) \quad \Sigma _{1}=E\left( \mathbf{x}_{1}\mathbf{x}%
_{1}^{\prime }\right) =W_{1},%
\end{array}%
\right.
\end{equation*}%
where $\widehat{\mathbf{e}}_{t}=\mathbf{y}_{t}-\widehat{\mathbf{y}}_{t}$ is
the $\mathbf{y}_{t}$-residuals with covariance matrix $\Omega _{t}$, $\Sigma
_{t}=E\left[ \left( \mathbf{x}_{t}-\widehat{\mathbf{x}}_{t}\right) \right. $
$\left. \left( \mathbf{x}_{t}-\widehat{\mathbf{x}}_{t}\right) ^{\prime }%
\right] $ is interpreted as the covariance matrix of the one-step state
prediction errors, and $K_{t}=E\left( \mathbf{x}_{t+1}\widehat{\mathbf{e}}%
_{t}^{\prime }\right) $\ is known as the Kalman gain. The notation $A\geq 0$
means that the matrix $A$ is nonnegative definite.

Recursion $(2e)$ based on the starting equation $(2g)$ will be called
\textit{periodic Riccati difference equation} $(PRDE)$ because in the limit,
i.e. when $\Sigma _{t+Sk}$ converges as $k\rightarrow \infty $ for all $t\in
\{1,...,S\}$, the $S$-periodic limiting solution $P_{t}=\underset{%
k\rightarrow \infty }{\lim }\Sigma _{t+Sk}$ will satisfy the following
\textit{discrete-time matrix periodic Riccati equation} $(DPRE)$%
\begin{equation*}
P_{t+1}=F_{t}P_{t}F_{t}^{\prime }-F_{t}P_{t}H_{t}\left( H_{t}^{\prime
}P_{t}H_{t}+R_{t}\right) ^{-1}H_{t}^{\prime }P_{t}F_{t}^{\prime
}+G_{t}Q_{t}G_{t}^{\prime },\text{ }t\in \{1,...,S\},\vspace{0.25cm}
\end{equation*}%
which has been extensively studied (see for example Bittanti et \textit{al},
$1988$ for some theoretical aspects and Hench and Laub, $1994$ for a
numerical resolution). As is well known, the resolution of $(2e)$ requires $%
O(r^{3})$ operations per iteration which is computationally expensive.
Furthermore, the solution $\Sigma _{t}$ must be nonnegative definite, a
property that is not easy to preserve in a numerical resolution of $(2e)$.
The following section proposes some recursions that avoid these drawbacks
and may have further advantages over the Kalman filter $(2)$.

\section{Periodic Chandrasekhar-type algorithms}

The recursions proposed in this section and which are aimed to generalize
Morf et \textit{al}'s $(1974)$ algorithms to the periodic case will be
called analogously periodic Chandrasekhar-type equations. This, of course,
will not mean that there is an analog of our recursions in the periodic
continuous-time case. The derivation of our recursions is similar to its
classical counterpart and is based on the factorization result given below
(see \textit{Theorem 3.1}).

Let $\Delta _{S}\Sigma _{t}=\Sigma _{t+S}-\Sigma _{t}$ denote the $S$-lagged
increment of the Riccati variable, for given $\Sigma _{1},\Sigma
_{2},...,\Sigma _{S}\geq 0$. Then, one can proves the following result.

\noindent \textbf{Theorem 3.1} \textit{The }$S$\textit{-lagged increment }$%
\Delta _{S}\Sigma _{t}$\textit{\ satisfies the following difference equations%
}%
\begin{equation}
\Delta _{S}\Sigma _{t+1}=\left( F_{t}-K_{t+S}\Omega _{t+S}^{-1}H_{t}^{\prime
}\right) \left[ \Delta _{S}\Sigma _{t}+\Delta _{S}\Sigma _{t}H_{t}\Omega
_{t}^{-1}H_{t}^{\prime }\Delta _{S}\Sigma _{t}\right] \left(
F_{t}-K_{t+S}\Omega _{t+S}^{-1}H_{t}^{\prime }\right) ^{\prime },  \tag{$3$}
\end{equation}%
\begin{equation}
\Delta _{S}\Sigma _{t+1}=\left( F_{t}-K_{t}\Omega _{t}^{-1}H_{t}^{\prime
}\right) \left[ \Delta _{S}\Sigma _{t}-\Delta _{S}\Sigma _{t}H_{t}\Omega
_{t+S}^{-1}H_{t}^{\prime }\Delta _{S}\Sigma _{t}\right] \left(
F_{t}-K_{t}\Omega _{t}^{-1}H_{t}^{\prime }\right) ^{\prime }.  \tag{$4$}
\end{equation}%
\noindent \textbf{Proof}

\noindent $i)$ \textbf{Proof of (3)}

From $(2a)$ we have%
\begin{eqnarray*}
\Omega _{t+S} &=&H_{t}^{^{\prime }}\Delta _{S}\Sigma
_{t}H_{t}+H_{t}^{^{\prime }}\Sigma _{t}H_{t}+R_{t} \\
&=&H_{t}^{^{\prime }}\Delta _{S}\Sigma _{t}H_{t}+\Omega _{t}.
\end{eqnarray*}%
Hence%
\begin{equation}
\Omega _{t+S}=\Omega _{t}+\Delta _{S}\Omega _{t},  \tag{$5$}
\end{equation}%
where $\Delta _{S}\Omega _{t}\overset{def}{=}H_{t}^{^{\prime }}\Delta
_{S}\Sigma _{t}H_{t}$. Moreover, from $(2e)$ it follows that%
\begin{eqnarray*}
\Sigma _{t+1+S} &=&F_{t}\Sigma _{t+S}F_{t}^{\prime }-\widetilde{K}%
_{t+S}\Omega _{t+S}\widetilde{K}_{t+S}^{^{\prime }}+G_{t}Q_{t}G_{t}^{\prime
}, \\
\Sigma _{t+1} &=&F_{t}\Sigma _{t}F_{t}^{\prime }-\widetilde{K}_{t}\Omega _{t}%
\widetilde{K}_{t}^{^{\prime }}+G_{t}Q_{t}G_{t}^{\prime },
\end{eqnarray*}%
where $\widetilde{K}_{t}=K_{t}\Omega _{t}^{-1}$. Therefore,%
\begin{equation}
\Delta _{S}\Sigma _{t+1}=F_{t}\Delta _{S}\Sigma _{t}F_{t}^{\prime }-%
\widetilde{K}_{t+S}\Omega _{t+S}\widetilde{K}_{t+S}^{^{\prime }}+\widetilde{K%
}_{t}\Omega _{t}\widetilde{K}_{t}^{^{\prime }}\text{.}  \tag{$6$}
\end{equation}%
On the other hand, the Kalman gain $\widetilde{K}_{t}$ may be written in a
backward recursive form as follows%
\begin{eqnarray*}
\widetilde{K}_{t} &=&\left( F_{t}\Sigma _{t+S}H_{t}-F_{t}\Delta _{S}\Sigma
_{t}H_{t}\right) \Omega _{t}^{-1} \\
&=&\left( \widetilde{K}_{t+S}\Omega _{t+S}-F_{t}\Delta _{S}\Sigma
_{t}H_{t}\right) \Omega _{t}^{-1} \\
&=&\left[ \widetilde{K}_{t+S}\left( H_{t}^{^{\prime }}\Sigma
_{t+S}H_{t}+R_{t}\right) -F_{t}\Delta _{S}\Sigma _{t}H_{t}\right] \Omega
_{t}^{-1} \\
&=&\left[ \widetilde{K}_{t+S}\left( H_{t}^{^{\prime }}\Delta _{S}\Sigma
_{t}H_{t}+H_{t}^{^{\prime }}\Sigma _{t}H_{t}+R_{t}\right) -F_{t}\Delta
_{S}\Sigma _{t}H_{t}\right] \Omega _{t}^{-1} \\
&=&\left[ \widetilde{K}_{t+S}\Omega _{t}+\widetilde{K}_{t+S}H_{t}^{^{\prime
}}\Delta _{S}\Sigma _{t}H_{t}-F_{t}\Delta _{S}\Sigma _{t}H_{t}\right] \Omega
_{t}^{-1} \\
&=&\widetilde{K}_{t+S}-\left( F_{t}-\widetilde{K}_{t+S}H_{t}^{^{\prime
}}\right) \Delta _{S}\Sigma _{t}H_{t}\Omega _{t}^{-1}.
\end{eqnarray*}%
Whence%
\begin{equation}
\widetilde{K}_{t}=\widetilde{K}_{t+S}-\Delta _{S}\widetilde{K}_{t},
\tag{$7$}
\end{equation}%
with $\Delta _{S}\widetilde{K}_{t}\overset{def}{=}\left( F_{t}-\widetilde{K}%
_{t+S}H_{t}^{^{\prime }}\right) \Delta _{S}\Sigma _{t}H_{t}\Omega _{t}^{-1}$.

Now replacing the latter expression of $\widetilde{K}_{t}$ in the last term
of the right hand side of $(6)$ while using $(5)$, we obtain%
\begin{eqnarray*}
\Delta _{S}\Sigma _{t+1} &=&F_{t}\Delta _{S}\Sigma _{t}F_{t}^{\prime }-%
\widetilde{K}_{t+S}\Omega _{t+S}\widetilde{K}_{t+S}^{^{\prime }}+\widetilde{K%
}_{t}\Omega _{t}\widetilde{K}_{t}^{^{\prime }} \\
&=&F_{t}\Delta _{S}\Sigma _{t}F_{t}^{\prime }-\widetilde{K}_{t+S}\left(
\Omega _{t}+H_{t}^{^{\prime }}\Delta _{S}\Sigma _{t}H_{t}\right) \widetilde{K%
}_{t+S}^{^{\prime }}+\left( \widetilde{K}_{t+S}-\Delta _{S}\widetilde{K}%
_{t}\right) \Omega _{t}\left( \widetilde{K}_{t+S}-\Delta _{S}\widetilde{K}%
_{t}\right) ^{^{\prime }} \\
&=&F_{t}\Delta _{S}\Sigma _{t}F_{t}^{\prime }-\widetilde{K}_{t+S}\Omega _{t}%
\widetilde{K}_{t+S}^{^{\prime }}-\widetilde{K}_{t+S}H_{t}^{^{\prime }}\Delta
_{S}\Sigma _{t}H_{t}\widetilde{K}_{t+S}^{^{\prime }} \\
&&+\widetilde{K}_{t+S}\Omega _{t}\widetilde{K}_{t+S}^{^{\prime }}-\widetilde{%
K}_{t+S}\Omega _{t}\Delta _{S}\widetilde{K}_{t}^{^{\prime }}-\Delta _{S}%
\widetilde{K}_{t}\Omega _{t}\widetilde{K}_{t+S}^{^{\prime }}+\Delta _{S}%
\widetilde{K}_{t}\Omega _{t}\Delta _{S}\widetilde{K}_{t}^{^{\prime }} \\
&=&F_{t}\Delta _{S}\Sigma _{t}F_{t}^{\prime }-\widetilde{K}%
_{t+S}H_{t}^{^{\prime }}\Delta _{S}\Sigma _{t}H_{t}\widetilde{K}%
_{t+S}^{^{\prime }}-\widetilde{K}_{t+S}H_{t}^{^{\prime }}\Delta _{S}\Sigma
_{t}\left( F_{t}-\widetilde{K}_{t+S}H_{t}^{^{\prime }}\right) ^{^{\prime }}
\\
&&-\left( F_{t}-\widetilde{K}_{t+S}H_{t}^{^{\prime }}\right) \Delta
_{S}\Sigma _{t}H_{t}\widetilde{K}_{t+S}^{^{\prime }}+\left( F_{t}-\widetilde{%
K}_{t+S}H_{t}^{^{\prime }}\right) \Delta _{S}\Sigma _{t}H_{t}\Omega
_{t}^{-1}H_{t}^{^{\prime }}\Delta _{S}\Sigma _{t}\left( F_{t}-\widetilde{K}%
_{t+S}H_{t}^{^{\prime }}\right) ^{^{\prime }} \\
&=&\left( F_{t}-\widetilde{K}_{t+S}H_{t}^{^{\prime }}\right) \Delta
_{S}\Sigma _{t}F_{t}^{\prime }-\left( F_{t}-\widetilde{K}_{t+S}H_{t}^{^{%
\prime }}\right) \Delta _{S}\Sigma _{t}H_{t}\widetilde{K}_{t+S}^{^{\prime }}
\\
&&+\left( F_{t}-\widetilde{K}_{t+S}H_{t}^{^{\prime }}\right) \Delta
_{S}\Sigma _{t}H_{t}\Omega _{t}^{-1}H_{t}^{^{\prime }}\Delta _{S}\Sigma
_{t}\left( F_{t}-\widetilde{K}_{t+S}H_{t}^{^{\prime }}\right) ^{^{\prime }}
\\
&=&\left( F_{t}-\widetilde{K}_{t+S}H_{t}^{^{\prime }}\right) \left[ \Delta
_{S}\Sigma _{t}+\Delta _{S}\Sigma _{t}H_{t}\Omega _{t}^{-1}H_{t}^{^{\prime
}}\Delta _{S}\Sigma _{t}\right] \left( F_{t}-\widetilde{K}%
_{t+S}H_{t}^{^{\prime }}\right) ^{^{\prime }},
\end{eqnarray*}%
proving $(3)$.

\noindent $ii)$ \textbf{Proof of (4)}

A similar argument may be used to prove $(4)$. It suffice to express $%
\widetilde{K}_{t+S}$ with respect of $\widetilde{K}_{t}$ in a forward
recursive form as follows%
\begin{eqnarray*}
\widetilde{K}_{t+S} &=&\left( F_{t}\Sigma _{t}H_{t}+F_{t}\Delta _{S}\Sigma
_{t}H_{t}\right) \Omega _{t+S}^{-1} \\
&=&\left( \widetilde{K}_{t}\Omega _{t}+F_{t}\Delta _{S}\Sigma
_{t}H_{t}\right) \Omega _{t+S}^{-1} \\
&=&\left[ \widetilde{K}_{t}\left( H_{t}^{^{\prime }}\Sigma
_{t}H_{t}+R_{t}\right) +F_{t}\Delta _{S}\Sigma _{t}H_{t}\right] \Omega
_{t+S}^{-1} \\
&=&\left[ \widetilde{K}_{t}\left( H_{t}^{^{\prime }}\Sigma
_{t+S}H_{t}-H_{t}^{^{\prime }}\Delta _{S}\Sigma _{t}H_{t}+R_{t}\right)
+F_{t}\Delta _{S}\Sigma _{t}H_{t}\right] \Omega _{t+S}^{-1} \\
&=&\left[ \widetilde{K}_{t}\Omega _{t+S}-\widetilde{K}_{t}H_{t}^{^{\prime
}}\Delta _{S}\Sigma _{t}H_{t}+F_{t}\Delta _{S}\Sigma _{t}H_{t}\right] \Omega
_{t+S}^{-1} \\
&=&\widetilde{K}_{t}+\left( F_{t}-\widetilde{K}_{t}H_{t}^{^{\prime }}\right)
\Delta _{S}\Sigma _{t}H_{t}\Omega _{t+S}^{-1},
\end{eqnarray*}%
that is%
\begin{equation}
\widetilde{K}_{t+S}=\widetilde{K}_{t}+\Delta _{S}\widetilde{K}_{t},\hspace{%
1cm}  \tag{$8$}
\end{equation}%
where $\Delta _{S}\widetilde{K}_{t}=\left( F_{t}-\widetilde{K}%
_{t}H_{t}^{^{\prime }}\right) \Delta _{S}\Sigma _{t}H_{t}\Omega _{t+S}^{-1}.$

Then, replacing the expression of $\widetilde{K}_{t+S}$ given by $(8)$ in
the second term of the right hand side of $(6)$, it follows that%
\begin{eqnarray*}
\Delta _{S}\Sigma _{t+1} &=&F_{t}\Delta _{S}\Sigma _{t}F_{t}^{\prime }-%
\widetilde{K}_{t+S}\Omega _{t+S}\widetilde{K}_{t+S}^{^{\prime }}+\widetilde{K%
}_{t}\Omega _{t}\widetilde{K}_{t}^{^{\prime }} \\
&=&F_{t}\Delta _{S}\Sigma _{t}F_{t}^{\prime }-\left( \widetilde{K}%
_{t}+\Delta _{S}\widetilde{K}_{t}\right) \Omega _{t+S}\left( \widetilde{K}%
_{t}+\Delta _{S}\widetilde{K}_{t}\right) ^{^{\prime }} \\
&&+\widetilde{K}_{t}\left( \Omega _{t+S}-H_{t}^{^{\prime }}\Delta _{S}\Sigma
_{t}H_{t}\right) \widetilde{K}_{t}^{^{\prime }} \\
&=&F_{t}\Delta _{S}\Sigma _{t}F_{t}^{\prime }-\widetilde{K}_{t}\Omega _{t+S}%
\widetilde{K}_{t}^{^{\prime }}-\widetilde{K}_{t}\Omega _{t+S}\Delta _{S}%
\widetilde{K}_{t}^{^{\prime }}-\Delta _{S}\widetilde{K}_{t}\Omega _{t+S}%
\widetilde{K}_{t}^{^{\prime }}-\Delta _{S}\widetilde{K}_{t}\Omega
_{t+S}\Delta _{S}\widetilde{K}_{t}^{^{\prime }} \\
&&+\widetilde{K}_{t}\Omega _{t+S}\widetilde{K}_{t}^{^{\prime }}-\widetilde{K}%
_{t}H_{t}^{^{\prime }}\Delta _{S}\Sigma _{t}H_{t}\widetilde{K}_{t}^{^{\prime
}} \\
&=&F_{t}\Delta _{S}\Sigma _{t}F_{t}^{\prime }-\widetilde{K}_{t}\Omega
_{t+S}\Delta _{S}\widetilde{K}_{t}^{^{\prime }}-\Delta _{S}\widetilde{K}%
_{t}\Omega _{t+S}\widetilde{K}_{t}^{^{\prime }}-\Delta _{S}\widetilde{K}%
_{t}\Omega _{t+S}\Delta _{S}\widetilde{K}_{t}^{^{\prime }}-\widetilde{K}%
_{t}H_{t}^{^{\prime }}\Delta _{S}\Sigma _{t}H_{t}\widetilde{K}_{t}^{^{\prime
}}.
\end{eqnarray*}%
Finally, using again $(8)$ we can write $\Delta _{S}\Sigma _{t+1}$ as follows%
\begin{eqnarray*}
\Delta _{S}\Sigma _{t+1} &=&F_{t}\Delta _{S}\Sigma _{t}F_{t}^{\prime }-%
\widetilde{K}_{t}\left( \left( F_{t}-\widetilde{K}_{t}H_{t}^{^{\prime
}}\right) \Delta _{S}\Sigma _{t}H_{t}\right) ^{^{\prime }}-\left( F_{t}-%
\widetilde{K}_{t}H_{t}^{^{\prime }}\right) \Delta _{S}\Sigma _{t}H_{t}%
\widetilde{K}_{t}^{^{\prime }} \\
&&-\left( \left( F_{t}-\widetilde{K}_{t}H_{t}^{^{\prime }}\right) \Delta
_{S}\Sigma _{t}H_{t}\right) \Omega _{t+S}^{-1}\left( \left( F_{t}-\widetilde{%
K}_{t}H_{t}^{^{\prime }}\right) \Delta _{S}\Sigma _{t}H_{t}\right)
^{^{\prime }}-\widetilde{K}_{t}H_{t}^{^{\prime }}\Delta _{S}\Sigma _{t}H_{t}%
\widetilde{K}_{t}^{^{\prime }} \\
&=&F_{t}\Delta _{S}\Sigma _{t}F_{t}^{\prime }-\widetilde{K}%
_{t}H_{t}^{^{\prime }}\Delta _{S}\Sigma _{t}\left( F_{t}-\widetilde{K}%
_{t}H_{t}^{^{\prime }}\right) ^{^{\prime }}-\left( F_{t}-\widetilde{K}%
_{t}H_{t}^{^{\prime }}\right) \Delta _{S}\Sigma _{t}H_{t}\widetilde{K}%
_{t}^{^{\prime }} \\
&&-\left( F_{t}-\widetilde{K}_{t}H_{t}^{^{\prime }}\right) \Delta _{S}\Sigma
_{t}H_{t}\Omega _{t+S}^{-1}H_{t}^{^{\prime }}\Delta _{S}\Sigma _{t}\left(
F_{t}-\widetilde{K}_{t}H_{t}^{^{\prime }}\right) ^{^{\prime }}-\widetilde{K}%
_{t}H_{t}^{^{\prime }}\Delta _{S}\Sigma _{t}H_{t}\widetilde{K}_{t}^{^{\prime
}} \\
&=&F_{t}\Delta _{S}\Sigma _{t}F_{t}^{\prime }-\widetilde{K}%
_{t}H_{t}^{^{\prime }}\Delta _{S}\Sigma _{t}F_{t}^{^{\prime }}+\widetilde{K}%
_{t}H_{t}^{^{\prime }}\Delta _{S}\Sigma _{t}\left( \widetilde{K}%
_{t}H_{t}^{^{\prime }}\right) ^{^{\prime }}-F_{t}\Delta _{S}\Sigma _{t}H_{t}%
\widetilde{K}_{t}^{^{\prime }} \\
&&-\left( F_{t}-\widetilde{K}_{t}H_{t}^{^{\prime }}\right) \Delta _{S}\Sigma
_{t}H_{t}\Omega _{t+S}^{-1}H_{t}^{^{\prime }}\Delta _{S}\Sigma _{t}\left(
F_{t}-\widetilde{K}_{t}H_{t}^{^{\prime }}\right) ^{^{\prime }} \\
&=&\left( F_{t}-\widetilde{K}_{t}H_{t}^{^{\prime }}\right) \Delta _{S}\Sigma
_{t}F_{t}^{^{\prime }}-\left( F_{t}-\widetilde{K}_{t}H_{t}^{^{\prime
}}\right) \Delta _{S}\Sigma _{t}\left( \widetilde{K}_{t}H_{t}^{^{\prime
}}\right) ^{^{\prime }} \\
&&-\left( F_{t}-\widetilde{K}_{t}H_{t}^{^{\prime }}\right) \Delta _{S}\Sigma
_{t}H_{t}\Omega _{t+S}^{-1}H_{t}^{^{\prime }}\Delta _{S}\Sigma _{t}\left(
F_{t}-\widetilde{K}_{t}H_{t}^{^{\prime }}\right) ^{^{\prime }} \\
&=&\left( F_{t}-\widetilde{K}_{t}H_{t}^{^{\prime }}\right) \Delta _{S}\Sigma
_{t}\left( F_{t}-\widetilde{K}_{t}H_{t}^{^{\prime }}\right) ^{^{\prime
}}-\left( F_{t}-\widetilde{K}_{t}H_{t}^{^{\prime }}\right) \Delta _{S}\Sigma
_{t}H_{t}\Omega _{t+S}^{-1}H_{t}^{^{\prime }}\Delta _{S}\Sigma _{t}\left(
F_{t}-\widetilde{K}_{t}H_{t}^{^{\prime }}\right) ^{^{\prime }},
\end{eqnarray*}%
showing $(4)$. $\blacksquare $

\textit{Theorem 3.1} shows that $\Delta _{S}\Sigma _{t}$ may be factorized
as follows%
\begin{equation}
\Delta _{S}\Sigma _{t}=Y_{t}M_{t}Y_{t}^{\prime },  \tag{$9$}
\end{equation}%
where $M_{t}$ is a square symmetric matrix, non necessarily nonnegative
definite, of dimension \newline
$rank\left( \Delta _{S}\Sigma _{1}\right) $, which is at least equal to $%
rank\left( \Delta _{S}\Sigma _{t}\right) $. Indeed, from $(3)$ we have%
\begin{equation*}
rank\left( \Delta _{S}\Sigma _{t+1}\right) \leq rank\left( \Delta _{S}\Sigma
_{t}\right) \leq ...\leq rank\left( \Delta _{S}\Sigma _{1}\right) \leq r.
\end{equation*}%
This can be exploited to derive some recursions with a best computational
complexity than the filter $(2)$.

Let us remark that \textit{Theorem 3.1} is not surprising since one can
always write a periodically time varying state-space model $(1)$ as a
time-invariant state space model (see Meyer and Burrus, $1975$) to which it
may be possible to apply the standard Chandrasekhar type factorization due
to Morf et \textit{al} $(1974)$. Nevertheless, because of the requiring
increasing bookkeeping (the obtained time invariant system is of dimension
multiplied by $S$) the development of a proper theory for periodic
state-space models would be fruitful.

Thanks to the factorization result given by \textit{Theorem 3.1}, the
matrices $Y_{t}$ and $M_{t}$ given by $(9)$ can be obtained recursively. The
following algorithm shows that the periodic Riccati difference equation $%
(2e) $ may be replaced by a set of recursions on $\Omega _{t}$, $K_{t}$, $%
Y_{t}$ and $M_{t}$ with a reduction in computational efforts, especially
when the state dimension $r$ is much larger than $m$, the dimension of $%
\mathbf{y}_{t} $.

\noindent \textbf{Algorithm 3.1} \textit{The Kalman filter }$(2)$\textit{\
can be replaced by a set of recursive equations containing }$(2c)$ and%
\textit{\ }$(2d)$\textit{\ and the following recursions}%
\begin{equation}
\left\{
\begin{array}{l}
\left( a\right) \quad \Omega _{t+S}=\Omega _{t}+H_{t}^{\prime
}Y_{t}M_{t}Y_{t}^{\prime }H_{t}, \\
\left( b\right) \quad K_{t+S}=\left( K_{t}+F_{t}Y_{t}M_{t}Y_{t}^{\prime
}H_{t}\right) , \\
\left( c\right) \quad Y_{t+1}=\left( F_{t}-K_{t+S}\Omega
_{t+S}^{-1}H_{t}^{\prime }\right) Y_{t}, \\
\left( d\right) \quad M_{t+1}=M_{t}+M_{t}Y_{t}^{\prime }H_{t}\Omega
_{t}^{-1}H_{t}^{\prime }Y_{t}M_{t},%
\end{array}%
\right.  \tag{$10$}
\end{equation}%
\textit{with starting values}%
\begin{equation*}
\left\{
\begin{array}{l}
\left( e\right) \quad \Omega _{s}=H_{s}^{\prime }\Sigma _{s}H_{s},\text{ \ }%
s=1,...,S, \\
\left( f\right) \quad K_{s}=F_{s}\Sigma _{s}H_{s},\text{ \ }s=1,...,S,%
\end{array}%
\right.
\end{equation*}%
\textit{where }$\Sigma _{s},$\textit{\ }$1\leq s\leq S$\textit{\ is
determined from} $(2e)$ \textit{and} $(2g)$, \textit{while }$Y_{1}$\textit{\
and }$M_{1}$\textit{\ are obtained by factorizing nonuniquely}
\begin{equation}
\Delta _{S}\Sigma _{1}=F_{S}\Sigma _{S}F_{S}^{\prime }-K_{S}\Omega
_{S}^{-1}K_{S}^{\prime }+G_{S}Q_{S}G_{S}^{\prime }-\Sigma _{1},  \tag{$10g$}
\end{equation}%
\textit{as}
\begin{equation*}
Y_{1}M_{1}Y_{1}^{\prime }.
\end{equation*}

\noindent \textbf{Derivation }$(10a)$ is just $(6)$\ when using $(9)$, while
$(10b)$ follows from $(9)$ and the relation
\begin{equation*}
K_{t+S}=\left( F_{t}\Sigma _{t}H_{t}+F_{t}\Delta _{S}\Sigma _{t}H_{t}\right)
.
\end{equation*}%
On the other hand, from $(3)$ which we rewrite while using $(9)$ we obtain
\begin{eqnarray*}
\Delta _{S}\Sigma _{t+1} &=&\left( F_{t}-\widetilde{K}_{t+S}H_{t}^{^{\prime
}}\right) \left[ \Delta _{S}\Sigma _{t}+\Delta _{S}\Sigma _{t}H_{t}\Omega
_{t}^{-1}H_{t}^{^{\prime }}\Delta _{S}\Sigma _{t}\right] \left( F_{t}-%
\widetilde{K}_{t+S}H_{t}^{^{\prime }}\right) ^{^{\prime }} \\
&=&\left( F_{t}-\widetilde{K}_{t+S}H_{t}^{^{\prime }}\right) \left[
Y_{t}M_{t}Y_{t}^{^{\prime }}+Y_{t}M_{t}Y_{t}^{^{\prime }}H_{t}\Omega
_{t}^{-1}H_{t}^{^{\prime }}Y_{t}M_{t}Y_{t}^{^{\prime }}\right] \left( F_{t}-%
\widetilde{K}_{t+S}H_{t}^{^{\prime }}\right) ^{^{\prime }} \\
&=&\left( F_{t}-K_{t+S}\Omega _{t+S}^{-1}H_{t}^{\prime }\right) Y_{t}\left(
M_{t}+M_{t}Y_{t}^{\prime }H_{t}\Omega _{t}^{-1}H_{t}^{\prime
}Y_{t}M_{t}\right) Y_{t}^{\prime }\left( F_{t}-K_{t+S}\Omega
_{t+S}^{-1}H_{t}^{\prime }\right) ^{\prime } \\
&=&Y_{t+1}M_{t+1}Y_{t+1}^{\prime }.
\end{eqnarray*}%
By simple identification we get $\left( 10c\right) $ and $\left( 10d\right) $%
. $\blacksquare $

Note that the $PRDE$ $(2e)$ must be executed for $1\leq s\leq S$ to start
recursions $(10)$. However, for $t>S$ the recursive calculation of $\Sigma
_{t}$ is not dealt with by the above algorithm but can be deduced from it
through the following equation%
\begin{equation*}
\Sigma _{kS+s}=\Sigma _{s}+\sum_{j=0}^{k-1}Y_{jS+s}M_{jS+s}Y_{jS+s}^{\prime
},\text{ \ }1\leq s\leq S.
\end{equation*}%
Similarly to the time-invariant case (Morf et \textit{al}, $1974$), other
forms of \textit{Algorithm 3.1} can be derived from \textit{Theorem 3.1}.
The following variant is particularly well adapted when $M_{1}<0$, in which
case we have $M_{t}\leq 0$ for any $t$. This case is encountered whenever
the periodic state-space model $(1)$ is periodically stationary (causal) as
we can see below.

\noindent \textbf{Algorithm 3.2} The following set of recursions in which $%
(10a)$, $(10b)$ and $(10e)$-$(10g)$ $(3.8a)$ are unchanged while $(10c)$ and
$(10d)$ are replaced by%
\begin{equation}
\left\{
\begin{array}{l}
\left( a\right) \quad Y_{t+1}=\left( F_{t}-K_{t}\Omega
_{t}^{-1}H_{t}^{\prime }\right) Y_{t}, \\
\left( b\right) \quad M_{t+1}=M_{t}-M_{t}Y_{t}^{\prime }H_{t}\Omega
_{t+S}^{-1}H_{t}^{\prime }Y_{t}M_{t},%
\end{array}%
\right.  \tag{$11$}
\end{equation}%
provides the same results as \textit{Algorithm 3.1}.

\noindent \textbf{Derivation} The derivation is similar to that of \textit{%
Algorithm 3.1}, but is based on the factorization $(4)$ rather than $(3)$. $%
\blacksquare $

It is still possible to derive other forms similarly to the standard
time-invariant case. The homogenous periodic Riccati difference equation $%
(10d)$ can be linearized using the matrix inversion lemma (Morf et \textit{al%
},\ $1974$) through which, we obtain a recursion on $M_{t}^{-1}$ rather than
on $M_{t}$ as follows%
\begin{equation*}
M_{t+1}^{-1}=M_{t}^{-1}-Y_{t}^{\prime }H_{t}\Omega _{t+S}^{-1}H_{t}^{\prime
}Y_{t}.
\end{equation*}%
It is worth noting that the periodic Chandrasekhar recursions given by
\textit{Algorithm 3.1} and \textit{Algorithm 3.2} will be preferred to the
Kalman filter $(2)$ whenever the dimension of $Y_{t}$ and/or $M_{t}$ are
significantly less than that of $\Sigma _{t}$. These dimensions are
conditioned on the good choice of the factorization $\Delta _{S}\Sigma
_{1}=Y_{1}M_{1}Y_{1}^{\prime }$ in the initialization step which will be
studied in the following section.

\section{The initialization problem}

As is well known, the most important step in the development of a
Chandrasekhar algorithm is the initialization step because it modulates the
computational complexity and hence the lack of numerical advantage over the
Kalman filter. In our periodic case, this step depends on the relation
between the period $S$, the output dimension $m$, and the state dimension $r$%
. First of all, suppose the process $\left\{ \mathbf{x}_{t}\right\} $ given
by $(1)$ is periodically stationary, that is, all the eingenvalues of the
monodromy matrix $\prod_{i=0}^{S}F_{S-i}$ are less than unity in modulus.
Let us consider two cases.

\noindent $i)$ \textbf{Case where} $Sm<r$:

As pointed out in $(10g)$ the start up values $Y_{1}$\textit{\ }and\textit{\
}$M_{1}$ are determined by factorizing $\Delta _{S}\Sigma _{1}$ as $%
Y_{1}M_{1}Y_{1}^{\prime }$. Iterating $(10g)$ $S$ times as follows%
\begin{eqnarray*}
\Delta _{S}\Sigma _{1} &=&F_{S}\Sigma _{S}F_{S}^{\prime }-\widetilde{K}%
_{S}\Omega _{S}\widetilde{K}_{S}^{^{\prime }}+G_{S}Q_{S}G_{S}^{\prime
}-\Sigma _{1} \\
&=&F_{S}\left( F_{S-1}\Sigma _{S-1}F_{S-1}^{\prime }-\widetilde{K}%
_{S-1}\Omega _{S-1}\widetilde{K}_{S-1}^{^{\prime
}}+G_{S-1}Q_{S-1}G_{S-1}^{\prime }\right) F_{S}^{\prime } \\
&&-\widetilde{K}_{S}\Omega _{S}\widetilde{K}_{S}^{^{\prime
}}+G_{S}Q_{S}G_{S}^{\prime }-\Sigma _{1} \\
&=&F_{S}F_{S-1}\Sigma _{S-1}\left( F_{S}F_{S-1}\right) ^{^{\prime }}-F_{S}%
\widetilde{K}_{S-1}\Omega _{S-1}\widetilde{K}_{S-1}^{^{\prime
}}F_{S}^{\prime }+F_{S}G_{S-1}Q_{S-1}G_{S-1}^{\prime }F_{S}^{\prime } \\
&&-\widetilde{K}_{S}\Omega _{S}\widetilde{K}_{S}^{^{\prime
}}+G_{S}Q_{S}G_{S}^{\prime }-\Sigma _{1}
\end{eqnarray*}%
\begin{eqnarray*}
&=&F_{S}F_{S-1}\left( F_{S-2}\Sigma _{S-2}F_{S-2}^{\prime }-\widetilde{K}%
_{S-2}\Omega _{S-2}\widetilde{K}_{S-2}^{^{\prime
}}+G_{S-2}Q_{S-2}G_{S-2}^{\prime }\right) \left( F_{S}F_{S-1}\right)
^{^{\prime }} \\
&&-F_{S}\widetilde{K}_{S-1}\Omega _{S-1}\widetilde{K}_{S-1}^{^{\prime
}}F_{S}^{\prime }+F_{S}G_{S-1}Q_{S-1}G_{S-1}^{\prime }F_{S}^{\prime }-%
\widetilde{K}_{S}\Omega _{S}\widetilde{K}_{S}^{^{\prime
}}+G_{S}Q_{S}G_{S}^{\prime }-\Sigma _{1} \\
&=&F_{S}F_{S-1}F_{S-2}\Sigma _{S-2}\left( F_{S}F_{S-1}F_{S-2}\right)
^{^{\prime }}-F_{S}F_{S-1}\widetilde{K}_{S-2}\Omega _{S-2}\left( F_{S}F_{S-1}%
\widetilde{K}_{S-2}\right) ^{^{\prime }} \\
&&+\left( F_{S}F_{S-1}\right) G_{S-2}Q_{S-2}G_{S-2}^{\prime }\left(
F_{S}F_{S-1}\right) ^{^{\prime }} \\
&&-F_{S}\widetilde{K}_{S-1}\Omega _{S-1}\left( \widetilde{K}%
_{S-1}F_{S}\right) ^{^{\prime }}+F_{S}G_{S-1}Q_{S-1}G_{S-1}^{\prime
}F_{S}^{\prime }-\widetilde{K}_{S}\Omega _{S}\widetilde{K}_{S}^{^{\prime
}}+G_{S}Q_{S}G_{S}^{\prime }-\Sigma _{1} \\
&&\vdots \\
&=&-\sum_{k=0}^{S-1}\left( \prod_{j=0}^{k-1}F_{S-j}\right) \widetilde{K}%
_{S-k}\Omega _{S-k}\widetilde{K}_{S-k}^{\prime }\left(
\prod_{j=0}^{k-1}F_{S-j}\right) ^{\prime }+\left(
\prod_{j=0}^{S-1}F_{S-j}\right) \Sigma _{1}\left(
\prod_{j=0}^{S-1}F_{S-j}\right) ^{\prime }
\end{eqnarray*}%
\begin{equation}
\hspace{4cm}+\sum_{k=0}^{S-1}\left( \prod_{j=0}^{k-1}F_{S-j}\right)
G_{S-k}Q_{S-k}G_{S-k}^{\prime }\left( \prod_{j=0}^{k-1}F_{S-j}\right)
^{\prime }-\Sigma _{1},  \tag{$12$}
\end{equation}%
and invoking the fact that under the periodic stationarity assumption, $%
\Sigma _{1}$\ satisfies the following discrete-time periodic Lyapunov
equation $(DPLE)$ (e.g. Bittanti et \textit{al}, $1988$; Varga,\ $1997$)
\begin{equation*}
\Sigma _{1}=\left( \prod_{j=0}^{S-1}F_{S-j}\right) \Sigma _{1}\left(
\prod_{j=0}^{S-1}F_{S-j}\right) ^{\prime }+\sum_{k=0}^{S-1}\left(
\prod_{j=0}^{k-1}F_{S-j}\right) G_{S-k}Q_{S-k}G_{S-k}^{\prime }\left(
\prod_{j=0}^{k-1}F_{S-j}\right) ^{\prime },
\end{equation*}%
we conclude that the sum of the last three terms of the right hand-side of $%
(12)$ is zero.

Whence%
\begin{eqnarray*}
\Delta _{S}\Sigma _{1} &=&-\sum_{k=0}^{S-1}\left(
\prod_{j=0}^{k-1}F_{S-j}\right) \widetilde{K}_{S-k}\Omega _{S-k}\widetilde{K}%
_{S-k}^{\prime }\left( \prod_{j=0}^{k-1}F_{S-j}\right) ^{\prime } \\
&=&-\sum_{k=0}^{S-1}\left( \prod_{j=0}^{k-1}F_{S-j}\right) K_{S-k}\Omega
_{S-k}^{-1}K_{S-k}^{\prime }\left( \prod_{j=0}^{k-1}F_{S-j}\right) ^{\prime }
\end{eqnarray*}%
\begin{equation}
=-L\left(
\begin{array}{ccc}
\Omega _{S}^{-1} & \cdots & 0 \\
\vdots & \ddots & \vdots \\
0 & \cdots & \Omega _{1}^{-1}%
\end{array}%
\right) L^{\prime }=Y_{1}M_{1}Y_{1}^{\prime },  \tag{$13$}
\end{equation}%
where $L$ is given by%
\begin{equation*}
L=\left[ K_{S},F_{S}K_{S-1},F_{S}F_{S-1}K_{S-2},...,%
\prod_{j=0}^{S-1}F_{S-j}K_{1}\right] \text{.}
\end{equation*}%
Clearly, with such a factorization the dimension of $M_{1}$ (and hence of $%
M_{t}$ for every $t$) is equal to $mS$ which is less than $r$, the dimension
of the Riccati matrix associated with the Kalman filter $(2)$. Indeed, when $%
Sm$ is fairly less than $r$, the nonhomogeneous $PRDE$ $(2e)$ may be
replaced by the homogenous $PRDE$ $(11b)$ which is of lower dimension. For
instance, for $m=1$, the complexity of solving $(10d)$ or $(11b)$ when using
$(13)$ as an initialization step is of order $O(Sr^{2})$ which is
computationally simple to solve compared to the $PRDE$ $(2e)$. It is still
possible to improve the computation of $(13)$ by alleviating the formation
of the sums of products in $L$ by using the periodic Schur decomposition
(Bojanczyk et \textit{al}, $1992$; Hench and Laub, $1994$).

\noindent $ii)$ \textbf{Case where} $Sm\geq r$:

In this case the latter factorization given by (13) would be inefficient
since the dimension of $M_{t}$ is greater than that of $\Sigma _{t}$. Thus
we have to search for another factorization. We have
\begin{eqnarray*}
\Sigma _{1} &=&E\left( x_{1}-\widehat{x}_{1}\right) \left( x_{1}-\widehat{x}%
_{1}\right) ^{^{\prime }}=E\left( x_{1}x_{1}^{^{\prime }}\right) \\
&=&E\left( F_{S}x_{0}+G_{S}w_{0}\right) \left( F_{S}x_{0}+G_{S}w_{0}\right)
^{^{\prime }} \\
&=&F_{S}E\left( x_{0}x_{0}^{^{\prime }}\right) F_{S}^{^{\prime
}}+G_{S}E\left( w_{0}w_{0}^{^{\prime }}\right) G_{S}^{^{\prime }} \\
&=&F_{S}W_{0}F_{S}^{^{\prime }}+G_{S}Q_{S}G_{S}^{^{\prime }}.
\end{eqnarray*}%
Therefore,%
\begin{eqnarray*}
\Delta _{S}\Sigma _{1} &=&Y_{1}M_{1}Y_{1}^{^{\prime }} \\
&=&F_{S}\Sigma _{S}F_{S}^{\prime }-\widetilde{K}_{S}\Omega _{S}\widetilde{K}%
_{S}^{^{\prime }}+G_{S}Q_{S}G_{S}^{\prime }-F_{S}W_{0}F_{S}^{^{\prime
}}-G_{S}Q_{S}G_{S}^{^{\prime }} \\
&=&F_{S}\left( \Sigma _{S}-W_{0}\right) F_{S}^{\prime }-\widetilde{K}%
_{S}\Omega _{S}\widetilde{K}_{S}^{^{\prime }} \\
&=&F_{S}\left( \Sigma _{S}-W_{0}\right) F_{S}^{\prime }-\left( F_{S}\Sigma
_{S}H_{S}\Omega _{S}^{-1}\right) \Omega _{S}\left( F_{S}\Sigma
_{S}H_{S}\Omega _{S}^{-1}\right) ^{^{\prime }} \\
&=&F_{S}\left[ \Sigma _{S}-W_{0}-\left( \Sigma _{S}H_{S}\right) \Omega
_{S}^{-1}\left( \Sigma _{S}H_{S}\right) ^{^{\prime }}\right] F_{S}^{\prime }%
\text{.}
\end{eqnarray*}%
This allows to identify $Y_{1}$\ and $M_{1}$ as follows%
\begin{equation}
\left\{
\begin{array}{l}
Y_{1}=F_{S} \\
M_{1}=\Sigma _{S}-W_{0}-\left( \Sigma _{S}H_{S}\right) \Omega
_{S}^{-1}\left( \Sigma _{S}H_{S}\right) ^{^{\prime }}\text{.}%
\end{array}%
\right.  \tag{$14$}
\end{equation}%
With such an initialization, the$\ PRDE$ $(11b)$ has the same dimension as
that of the $PRDE$ $(2e)$, and it seems that there is no reduction in the
computational cost compared to the Kalman filter. However, the difference
from $(2e)$ is that, unlike the $\Sigma _{t}$, the $M_{t}$ is not required
to be nonnegative-definite. This helps alleviate the computational
complexity of $(10d)$ and then $(11b)$.

In the matter of illustration we propose the following example which shows
the impact of a good choice of a starting factorization on the Chandrasekhar
algorithm complexity.

\noindent \textbf{Example 4.1 }Consider\textbf{\ }a periodic autoregression
of order 5 and period $S$ ($PAR_{S}(5)$), which is given by the following
stochastic difference equation%
\begin{equation}
y_{t}-\phi _{1}^{\left( t\right) }y_{t-1}-\phi _{2}^{\left( t\right)
}y_{t-2}-\phi _{3}^{\left( t\right) }y_{t-3}-\phi _{4}^{\left( t\right)
}y_{t-4}-\phi _{5}^{\left( t\right) }y_{t-5}=\varepsilon _{t},  \tag{$15$}
\end{equation}%
where $\left\{ \varepsilon _{t}\right\} $ is a periodic white noise with $S$%
-periodic variance and where the parameters $\phi _{j}^{\left( t\right) }$, $%
j=1,...,5$ are periodic with respect of $t$ with $S$.

Setting $\mathbf{x}_{t}=(y_{t},y_{t-1}...,y_{t-4})^{\prime }$, $\mathbf{%
\varepsilon }_{t}=(\varepsilon _{t},0,0,0,0)^{\prime }$ and $H^{\prime
}=(1,0,0,0,0)^{\prime }$, model $(15)$ may be written in the state-space form%
\begin{equation}
\begin{array}{l}
\mathbf{x}_{t}=F_{t}\mathbf{x}_{t-1}+\mathbf{\varepsilon }_{t} \\
y_{t}=H^{\prime }\mathbf{x}_{t}%
\end{array}
\tag{$16$}
\end{equation}%
so that identifying it with model $(1)$, the dimensions $r$, $m$ and $d$ are
respectively equal to $5$, $1$ and $5$.

When applying the Kalman filter to model $(16)$, the corresponding periodic
Riccati equation $(2e)$ is of dimension $5$ (the dimension of $\Sigma _{t}$)
for any value of $S$. However, the dimension of the Riccati equation
corresponding to the periodic Chandrasekhar filter (dimension of $M_{t}$)
depends upon $S$.

Let us consider two cases for $S$.\newpage

\noindent $i)$ \textbf{Case where }$S=2$.

We are in the case where $Sm<r$. According to formula $(13)$, we have%
\begin{equation*}
\Delta _{S}\Sigma _{1}=-L\left(
\begin{array}{cc}
\Omega _{2}^{-1} & 0 \\
0 & \Omega _{1}^{-1}%
\end{array}%
\right) L^{\prime }=Y_{1}M_{1}Y_{1}^{\prime },
\end{equation*}%
with $L=\left[ K_{2},F_{2}K_{1}\right] $, $K_{1}=F_{1}\Sigma _{1}H_{1}$ and $%
\Omega _{t}=H_{t}^{\prime }\Sigma _{t}H_{t}$, $t=1,2$. So, we can take $%
M_{1}=\left(
\begin{array}{cc}
\frac{1}{\Omega _{2}} & 0 \\
0 & \frac{1}{\Omega _{1}}%
\end{array}%
\right) $, from which the corresponding Riccati equation is of dimension $2$%
, clearly lower than the dimension of the Riccati equation of the Kalman
filter. Whence in this case the periodic Chandrasekhar filter is highly
superior to its homologue, the Kalman one.

\noindent $ii)$ \textbf{Case where }$S=12$.

In this case, the previous factorization is inefficient since the dimension
of the Chandrasekhar Riccati would be equal to 12, much larger than 5, the
dimension of the Kalman Riccati. Nevertheless, we are in the case $Sm>r$,
and according to $(14)$, $\Delta _{S}\Sigma _{1}$ may be factorized as $%
Y_{1}M_{1}Y_{1}^{\prime }$, where%
\begin{equation*}
Y_{1}=F_{12}\text{ and }M_{1}=\Sigma _{12}-W_{0}-\Sigma _{12}H_{12}\Omega
_{12}^{-1}H_{12}^{\prime }\Sigma _{12}^{\prime },
\end{equation*}%
so that the Riccati equation associated with the Chandrasekhar filter has
the same dimension as that of the Riccati equation of the Kalman filter.
Moreover, the matrix $M_{1}$ is not necessarily nonnegative definite in
contrast with $\Sigma _{1}$, and from this viewpoint the Chandrasekhar
filter is still more suitable.

\section{Conclusion}

In this paper the discrete-time Chandrasekhar recursions have been
generalized to the periodic time-varying state-space case through several
forms. These recursions allow in a large range of cases to solve the
periodic Riccati difference equation with a considerable reduction in the
computational complexity. Along similar lines to the standard time-invariant
case (Morf and Kailath, $1975$), a square root version of these recursions
may be easily derived in order to improve the numerical stability of the
proposed algorithms. Useful applications for time series analysis as well as
for the periodic system theory can be given, in particular, we mention the
likelihood evaluation of periodic $VARMA$ (Aknouche and Hamdi, $2007$), the
calculation of exact Fisher information matrix for $PARMA$ models and the
development of fast $RLS$ algorithms for periodic systems (Bentarzi and
Aknouche, $2006$).

\textbf{Acknowledgements}\ The authors are deeply grateful to an anonymous
referee for his judicious suggestions that have considerably improved the
quality of the paper.

\end{document}